\documentclass[
 reprint,
 groupedaddress,
 showpacs,
 superscriptaddress, 
 amsmath,amssymb,
 aps,
 pra
]{revtex4-1}

\usepackage[pdftex]{graphicx, color}% Include figure files
\usepackage{multirow,booktabs}
\usepackage{bm}% bold math
\usepackage{braket}
\usepackage{siunitx}
\usepackage{physics}
\usepackage{hyperref}
\usepackage{ulem}
\usepackage{comment}
\usepackage{here}

\newcommand{\vv}{ {\bm v }}
\newcommand{\rr}{{\bm r}}
\newcommand{\kk}{{\bm k}}

\newcommand{\tA}{\tilde{A}}

\newcommand{\beb}{\begin{itembox}}
\newcommand{\enb}{\end{itembox}}

\newcommand{\E}{{\bm E}}
\newcommand{\B}{{\bm B}}

\newcommand{\unit}{\hat{{\bm e}}}

\begin{document}

\title{Nonlocal optical responses of ultrapure metals in the hydrodynamic regime}

\author{Riki Toshio}
\email[]{toshio.riki.63c@st.kyoto-u.ac.jp}
\affiliation{%
 Department of Physics, Kyoto University, Kyoto 606-8502, Japan
}%

\author{Kazuaki Takasan}
%\email[]{yanase@scphys.kyoto-u.ac.jp}
\affiliation{%
  Department of Physics, University of California, Berkeley, California 94720, USA
}%

\author{Norio Kawakami}
%\email[]{yanase@scphys.kyoto-u.ac.jp}
\affiliation{%
 Department of Physics, Kyoto University, Kyoto 606-8502, Japan
}%

%\altaffiliation[Also at ]{Physics Department, Kyoto University.}
%\homepage[]{Your web page}

\date{\today}

\begin{abstract}

Our conventional understanding of optical responses in metals has been based on the Drude theory.
In recent years, however, it has become possible to prepare ultrapure metallic samples where the electron-electron scattering becomes the most dominant process governing transport and thus the Drude theory, where momentum-relaxing scatterings such as electron-impurity scattering are assumed to be dominant, is no longer valid. 
This regime is called the hydrodynamic regime and described by an emergent hydrodynamic theory. 
Here, we develop a basic framework of optical responses in the hydrodynamic regime.
Based on the hydrodynamic equation, we reveal the existence of a ``hydrodynamic mode'' resulting from the viscosity effect and compute the reflectance and the transmittance in three-dimensional electron fluids. 
Our theory also describes how to probe the hydrodynamic effects and measure the viscosity through simple optical techniques.

\end{abstract}

\pacs{74.20.-z, 74.70.-b}

\maketitle

%%%%%%%%%% Introduction %%%%%%%%%%

%%自己流

\section{Introduction\label{introduction}}
Hydrodynamics is a general framework to describe the low-energy dynamics in interacting many-particle systems, and it is based on the following two assumptions~\cite{Landau,Chaikin}. The first is {\it the local quasi-equilibrium of fluids}, which is achieved only when an external perturbation varies slowly in space and time, relative to the electron-electron scattering length $l_{ee}$ and time $\tau_{ee}$. In other words, this means that the characteristic length and time scale of the dynamics,  $L$ and $T$, are much larger than those of scattering; $l_{ee}\ll L,\ \tau_{ee}\ll T$. 
The second is {\it the conservation of mass, momentum and energy of particles}. In electron fluids in metals, however,  this assumption does not hold due to electron-impurity, electron-phonon and umklapp scattering processes. 
Therefore, to apply hydrodynamics to the electrons in solids, we need to satisfy the condition $l_{ee}\ll l_{mr}$, $\tau_{ee} \ll \tau_{mr}$, where $\l_{mr}$ and $\ \tau_{mr}$ are  the mean free path and time of momentum-relaxing scattering. In reality, this condition is almost always unsatisfied and thus the transport in macroscopic scale is governed by momentum-relaxation processes. This regime is called the Drude regime where the electron dynamics is described by the Drude theory~\cite{Ashcroft}. 

In recent years, however, it has become possible to prepare ultrapure samples which satisfy the above condition over a certain temperature range~\cite{Molenkamp1994, Molenkamp1995,Moll2016, Gooth2018,Bandurin2016, Kumar2017}. In these systems, the dynamics is described by hydrodynamics and the viscosity of the fluid plays an important role in  transport and optical phenomena~\cite{Lucas2018,Gurzhi1968, Andreev2011, Mendoza2011, Tomadin2014, Iacopo2015, Alekseev2016,Thomas2017,Guo2017}. In fact, many pieces of evidence for hydrodynamic electron flow has already been  reported, through several DC transport phenomena, in ultrapure materials such as GaAs quantum wells~\cite{Molenkamp1994,Molenkamp1995}, two-dimesional (2D) monovalent layered metal PdCoO${}_2$~\cite{Moll2016}, Weyl semimetal WP${}_2$~\cite{Gooth2018}, and graphene~\cite{Bandurin2016, Kumar2017}.

In more recent years, the high-frequency flow and optical responses of viscous electron fluids began to be discussed theoretically~\cite{Alekseev2018resonance, Alekseev2018honey, Moessner2018, Iacopo2018, Sun2018, Semenyakin2018, Cohen2018, Svintsov2018}. In the hydrodynamic regime, the optical responses of electrons are affected by the nonlocality (viscosity) and the nonlinearity of electron fluids, giving rise to crucial differences from those in the Drude regime. 
For example, in Refs.~\cite{Alekseev2018resonance, Alekseev2018honey}, the ac flow of the 2D fluid in a magnetic field was studied and a novel resonant phenomenon, viscous\ resonance, was proposed, which manifests itself %itself in the damping coefficient of magneto-plasmons 
at a frequency $\omega$ equal to the doubled electron cyclotron frequency $2\omega_c$. 
In regard to nonlinear optical responses, in Ref.~\cite{Sun2018}, second-harmonic generation has been discussed in a Dirac fluid, which is a class of electron fluids realized in systems with a Dirac-like dispersion, such as graphene~\cite{Lucas2018}.
These previous studies have focused on 2D electron fluids, but hydrodynamic optical responses in 3D bulk materials have not been well examined yet, although most of real materials, including hydrodynamic materials such as PdCoO${}_2$ and WP${}_2$, are 3D ones. In these systems, we need to solve the hydrodynamic equation consistently with the Maxwell equations in 3D space to reveal the fundamental optical properties, such as reflectance and transmittance, while their electron properties have strong two dimensionality.

Furthermore, we are interested in optical phenomena of electron fluids not only for theoretical reasons, but also for practical reasons. This is because these optical phenomena may provide a more efficient probe of hydrodynamic effects, compared to DC transport phenomena. To detect hydrodynamic effects through DC transport, we need to prepare microfabricated samples to cause a spatial variation of the velocity by the size effect, requiring advanced microfabrication techniques and considerable labor. On the other hand, in optical responses, such a variation is caused by an electromagnetic wave in its wavelength scale and we have no need to process samples at the mesoscale level. For these reasons, it is expected that we can more easily detect hydrodynamic effects or measure the viscosity of electron fluids through simple optical techniques. 

In this paper, we study the optical responses of viscous electronic fluids inhabiting in 3D space which is described by the hydrodynamic equation. 
%Electron fluids with linear dispersion can also be described by the same equation within linear response regime in Fermi liquid limit $\mu\gg k_BT$~\cite{Lucas2018}.
We note that the following analysis is restricted to the 2D dynamics of electron fluids and so it is applicable to layered metals, such as PdCoO${}_2$.
Solving the hydrodynamic equation and the Maxwell equation consistently, we obtain the optical conductivity of electron fluids and the dispersion relations of electromagnetic waves, suggesting that there exist two propagating transverse modes in 3D electron fluids. Furthermore, we clarify the contribution of these modes to the reflectance and the transmittance of 3D electron fluids, which provide an optical probe of hydrodynamic effects. Finally, we address the possibility of second-order optical responses, such as the second harmonic generation. These effects result from the nonlinearlity of electron fluids and the multi-branch structure of electron fluids.

This paper is organized as follows. 
In Sec.~\ref{2}, we start with the calculation of the optical conductivity and the reflectance spectrum of 3D electron fluids. Here we show that, in the hydrodynamic regime, there are two propagation modes of transverse electromagnetic waves due to the nonlocality of electron fluids. These dispersion relations, as shown in Sec.~\ref{3}, give rise to a change in the reflectance spectrum of ultrapure metals, compared with the estimation by the Drude theory. In Sec.~\ref{4}, we calculate the transmittance spectrum of electromagnetic waves through a thin ultrapure metal. In the hydrodynamic regime, electromagnetic waves can penetrate more deeply than they do in the Drude regime and thus we can obtain much larger transmittance. Furthermore, we also find that the spectrum shows a characteristic peak in the THz frequency regime.  Finally, in Sec.~\ref{5}, we briefly summarize our results and discuss the possibility of the nonlocal second-order optical responses in 3D electron fluids, which result from the nonlinearity and multi-branch structure of electron fluids even in centrosymmetric crystals.

\begin{comment}
In order to propose an observable hydrodynamic effect, 
\end{comment}

%%%%%%%%%% 3D reflectivity %%%%%%%%%%

\section{Optical conductivity and dispersion relations\label{2}}

In this section, we consider the optical conductivity and reflectance of 3D electron fluids within linear response. The in-plane dynamics of 3D electron fluids is described by the following hydrodynamic equation~\cite{Landau, Alekseev2016}  
\begin{equation}\label{general}
\begin{split}
		\frac{\partial \vv}{\partial t} +(\vv\cdot \grad)\vv = -\frac1\rho \grad p
	  +\nu {\bm \Delta}\vv + \gamma \grad(\div\vv)\\
    +\frac{e\E}{m}+\left[(\nu_{h}\Delta \vv +\omega_c \vv)\times {\bm B}
		\right]-\frac{\vv}{\tau},
\end{split}
\end{equation}
where $\E(\rr,t)$ and $\B(\rr,t)$ are electromagnetic ac fields,
\begin{equation}\label{E}
\E(\rr,t) = \Re\left[\tilde{\E}(\kk,\omega) e^{i\kk\cdot \rr-i\omega t} \right] ,
\end{equation}
\begin{equation}
\B(\rr,t) = \Re\left[\tilde{\B}(\kk,\omega) e^{i\kk\cdot \rr-i\omega t} \right] ,
\end{equation}
and $\vv$, $\rho$, $p$, $\nu$ and $\nu_h$ denote, respectively, the velocity field, the mass density, the pressure, the kinematic viscosity, and the Hall viscosity of 3D electron fluids. $m$, $e$ and $\omega_c$ are the electron mass, charge and the cyclotron frequency and $\tau$ is the relaxation time for momentum relaxing scattering. 
Furthermore, using the spacial dimension $d$ and the bulk viscosity $\zeta$, we define the parameter $\gamma$ as follows:
\[
\gamma \equiv \left(\frac\zeta\rho +\frac{d-2}{d}\nu \right).
\]
If we regard the metal as an isotropic 3D electron fluid, we set $d=3$, on the other hand, if we discuss a layered metal such as PdCoO${}_2$, we set $d=2$. 
For a Fermi liquid, the bulk viscosity $\zeta$ is known to be relatively small: $\zeta \sim (T/\epsilon_F)^2 \eta$~\cite{Abrikosov1959}, where $T$ is temperature and $\epsilon_F$ is the Fermi energy. In this regard, we will neglect the bulk viscosity in the following analysis. Moreover, we approximate the pressure gradient term in Eq.~(\ref{general}) within linear response as follows:
\[
\frac{1}{\rho}\grad p = \frac{K}{\rho}  \frac{\grad n}{n} \simeq \frac{K}{mn_0^2}  \grad n,
\]
where $K$ denotes the bulk modulus and is defined by 
\[
K\equiv -V \left(\pdv{p}{V}\right)_{N,T} = n \left(\pdv{p}{n}\right)_{T}.
\]
Here we have assumed that we can neglect the effect of the temperature modulation resulting from the dissipation in electron dynamics. 
Under these approximations, we can obtain a steady-state solution of the form
\begin{equation}\label{v}
\vv(\rr,t) =\Re\left[ \tilde{\vv}(\kk,\omega)e^{i\kk\cdot \rr-i\omega t}\right] .
\end{equation}
Substituting Eqs.~(\ref{E}) and (\ref{v}) into Eq.~(\ref{general}), we find the following relation within linear response to $\E$:
\begin{eqnarray}\label{1}
\left[-i\omega+\nu \tilde{\kk}^2+1/\tau+\left(
\frac {iK}{mn_0\omega} +\gamma
\right)(\kk\kk)  \right]\tilde{\vv}(\kk,\omega) \nonumber  \\
= -\frac em\tilde{\E}(\kk,\omega),
\end{eqnarray}
where $\kk\kk$ denotes the dyadic product of $\kk$. 
As a consequence, we arrive at the optical conductivity of electron fluids as follows:

\begin{equation}\label{conductivity}
\tilde{j}_i(\kk,\omega) = en\tilde{\vv} (\kk,\omega)=\sigma_{ij}(\kk,\omega) \tilde{E}_j(\kk,\omega),
\end{equation}
\[
\sigma_{ij}(\kk,\omega) =\sigma_\perp(\kk,\omega)\left(\delta_{ij}-\frac{k_ik_j}{\kk^2} \right) +\sigma_\parallel(\kk,\omega) \frac{k_ik_j}{\kk^2},
\]
\begin{equation}\label{perp}
\sigma_\perp(\kk,\omega)= \frac{\sigma_0}{1-i\omega \tau +\nu \tau \kk^2},
\end{equation}
\begin{equation}\label{parallel}
\sigma_\parallel(\kk,\omega)= \frac{\sigma_0}{1-i\omega \tau+\left(
\frac {iK}{mn_0\omega} +\nu +\gamma
\right)\tau \kk^2 },
\end{equation}
where $\sigma_0=ne^2\tau/m$ is the Drude conductivity.  
We note that the $\kk$-dependence of the optical conductivity reflects the nonlocal propeties of electron fluids, that is, viscosity effects. This result is in contrast to the optical conductivity in the Drude theory $\sigma(\omega)=\sigma_0/(1-i\omega \tau)$, which does not have $\kk$-dependence~\cite{Ashcroft}.

Next, we determine the dispersion relations of electromagnetic waves in electron fluids. This is achieved by substituting Eq.~(\ref{conductivity}) into Maxwell equation
\begin{equation}
-\Delta \tilde{\E} + \grad(\div \tilde{\E})= \frac{i\omega}{c}\left(
\frac{4\pi}{c}\tilde{{\bm j}} +\frac1c \pdv{\tilde{\E}}{t}
\right).
\end{equation}
As a result, we obtain the following equations to determine the dispersion relations:
\begin{equation}\label{former}
1+\frac{4\pi i \sigma_\parallel(\kk,\omega)}{\omega} = 0,
\end{equation}
\begin{equation} \label{latter}
\kk^2 
=\frac{\omega^2 }{c^2}\left[1+
\frac{4\pi i \sigma_\perp(\kk,\omega)}{\omega}\right].
%=\frac{\omega^2 }{c^2}\left[1+
%\frac{4\pi i}{\omega}\frac{\sigma_0}{1-i\omega\tau +\nu \tau k^2}\right],
\end{equation}
The former~(\ref{former}) gives us the dispersion relation of the longitudinal mode, i.e. the plasma mode. Substituting~(\ref{parallel})  and solving the equation for $\omega$, in the clean limit($\tau \to \infty$), we arrive at a familiar form of the dispersion relation for the plasma mode~\cite{Fetter} as follows: 
\[
	\omega =-\frac{i(\nu+\gamma)k^2}{2} \pm \omega_p \sqrt{1+\frac{K}{mn_0\omega_p^2}k^2 - \left(
	\frac{\nu+\gamma}{\omega_p}
	\right)^2 k^4}
\]\[
=\pm \omega_p\left(
	1+\frac{K}{2mn_0\omega_p^2}k^2
	\right)-\frac{i(\nu+\gamma)k^2}{2}  +\mathcal{O}(k^4),
\]
where $\omega_p =  \sqrt{4\pi ne^2/m}$ is the plasma frequency. This means that, in this limit, the lifetime of plasma is determined by the viscosity coefficients. 

\begin{figure*}[t]
  \centering
  \begin{tabular}{ll}
(a)  &(b) \\
   \includegraphics[width=7.5cm]{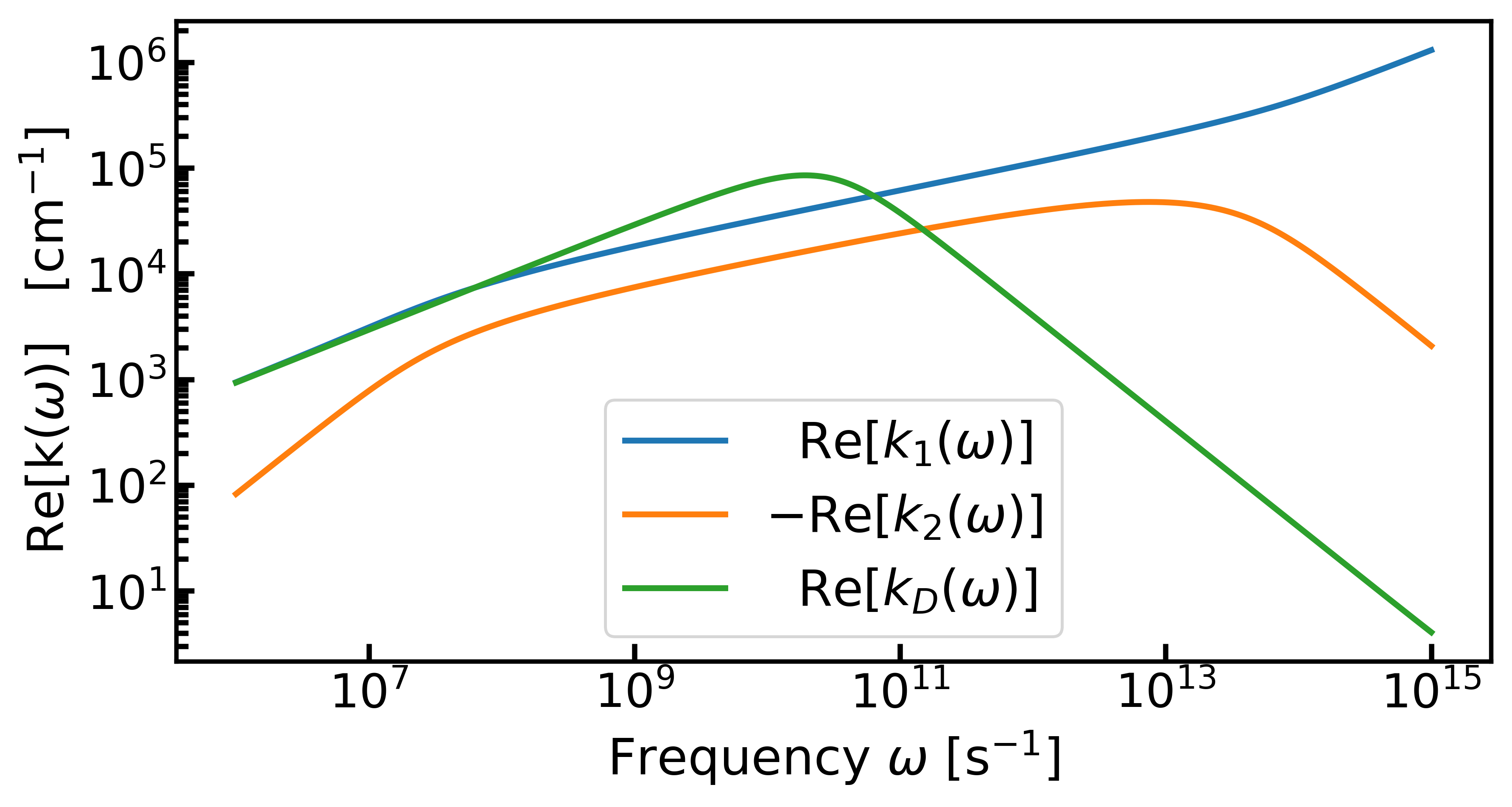}&
    \includegraphics[width=7.5cm]{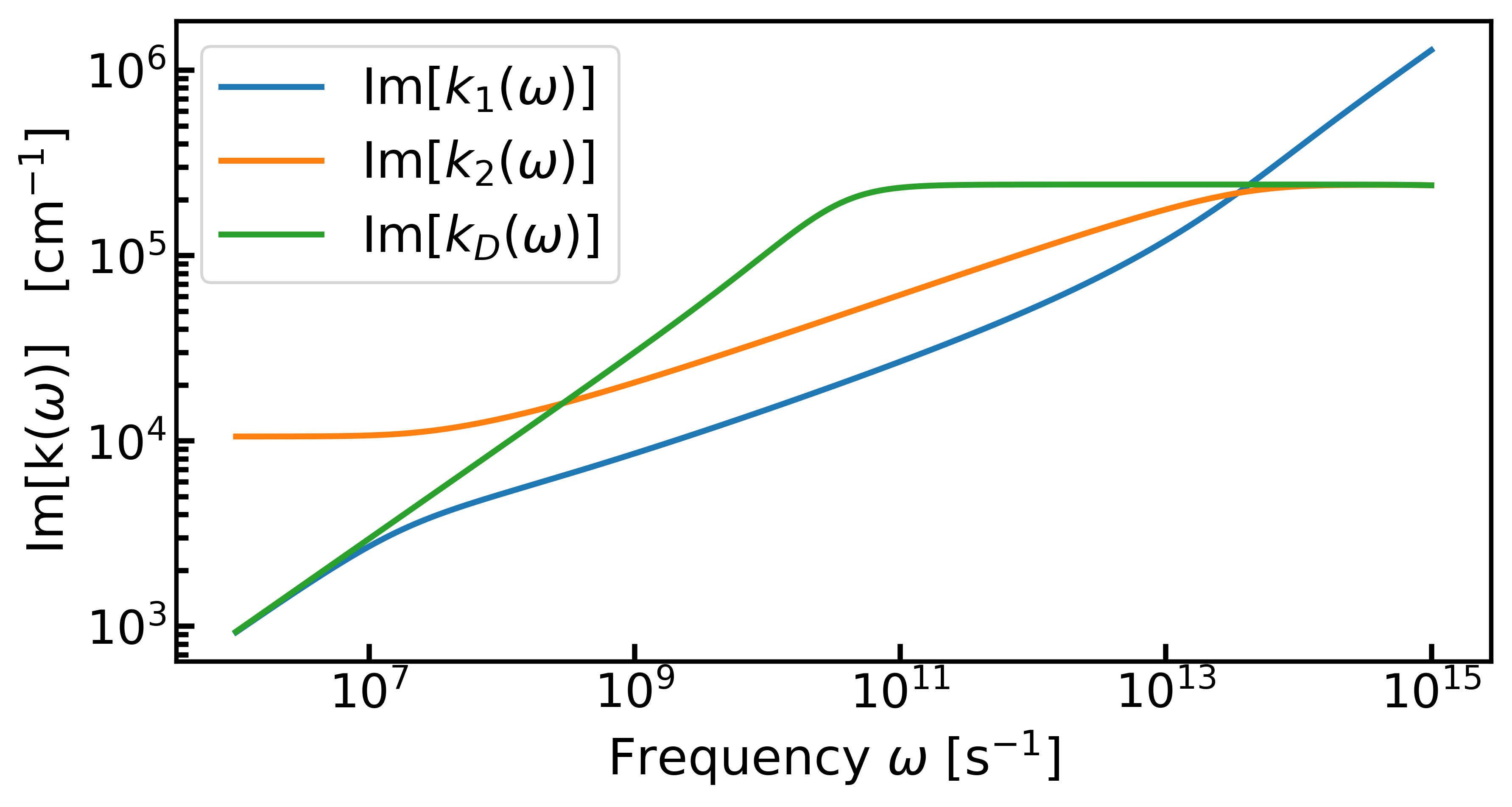}\\
(c) & (d) \\
   \includegraphics[width=7.5cm]{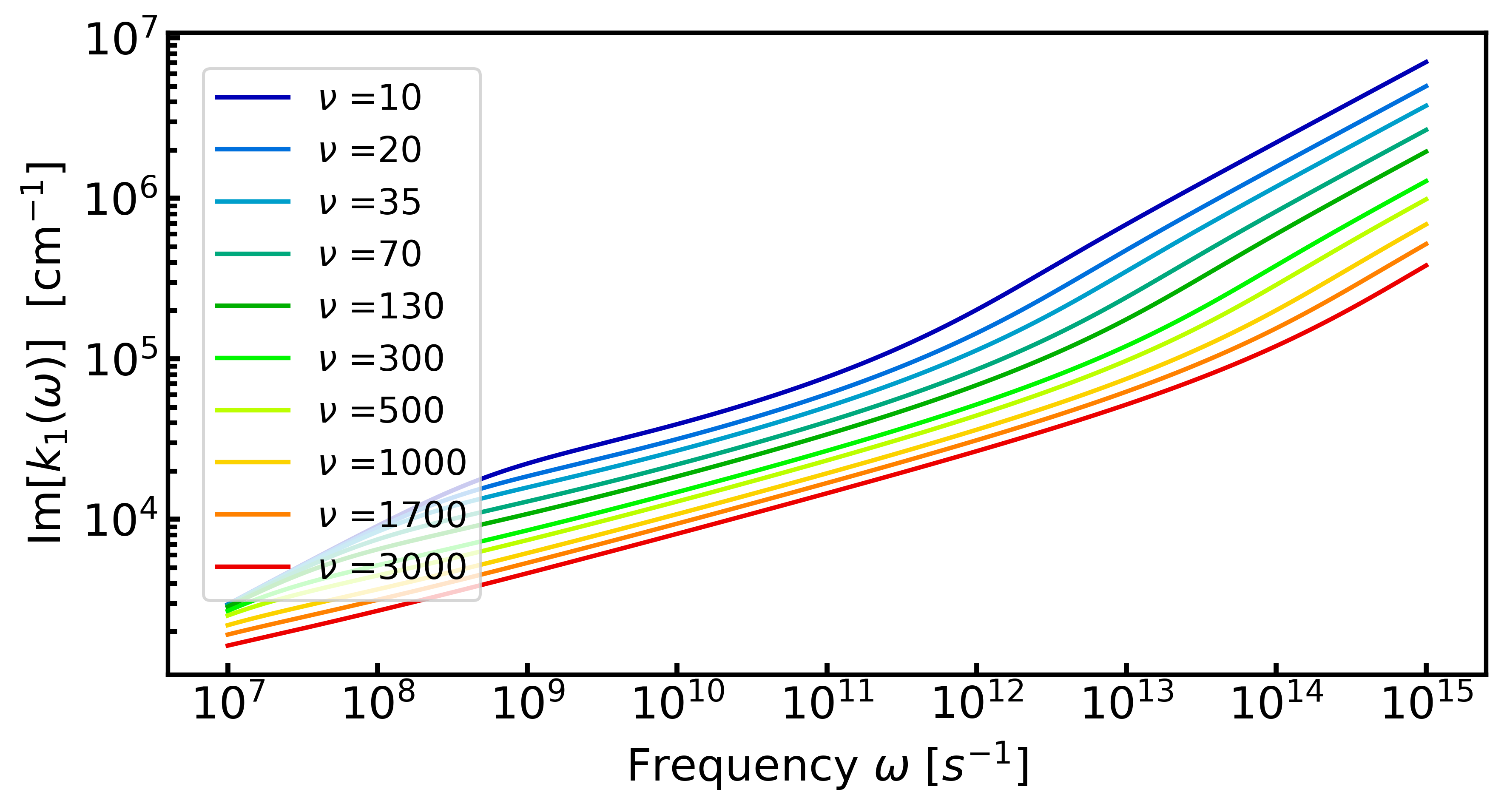}&
    \includegraphics[width=7.5cm]{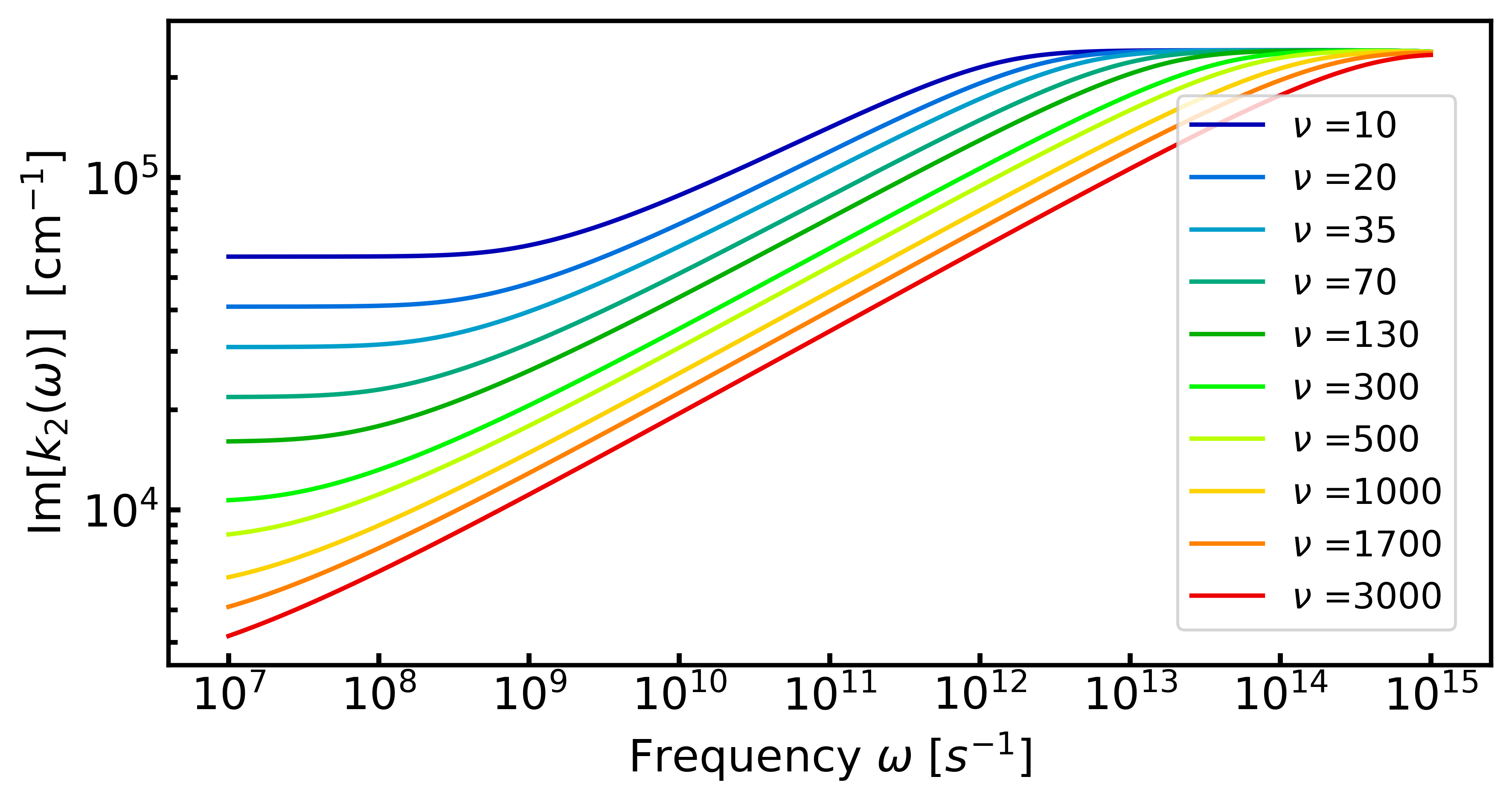}
  \end{tabular}
\caption{(a)(b) Dispersion relations $k_1(\omega)$ (blue) and $k_2(\omega)$ (yellow) estimated for a hydrodynamic material PdCoO${}_2$: For comparison, we also show the dispersion relation $k_D(\omega)$ deduced from the Drude theory (green).  Here, we define $k_i$ as the square root of $\kk_i^2$ which is chosen to make the imaginary part positive. We can see that $k_1(\omega)$ approaches $k_D(\omega)$ asymptotically in the low-frequency limit.  (c)(d) The viscosity-dependence of Im$[k_{1,2}(\omega)]$. Im$[k_{1,2}(\omega)]$ decreases monotonically as the viscosity increases.}
	\label{dispersion}
\end{figure*}

 On the other hand, Eq.(\ref{latter}) gives us the dispersion relations of the transverse modes. As can be seen readily by substituting Eq.(\ref{perp}), Eq.(\ref{latter}) leads to a quadratic equation of $\kk^2$ and the solutions read,
\begin{equation}\label{bunsan}
		\kk_{1,2}^2(\omega)=\frac{k_0^2}{2} \left[
		1-\alpha \pm \sqrt{(\alpha-1)^2 + 4\left(\alpha+\frac\beta\xi\right) }
		\right],
\end{equation}
where, in order to simplify the equation, we have introduced dimensionless parameters  
\[
k_0\equiv\frac\omega c, \  \xi\equiv k_0^2\nu\tau,\ \alpha\equiv \frac{1-i\omega\tau}{\xi},\ \beta\equiv \frac{4\pi i\sigma_0}{\omega}=\frac{i\omega_p^2 \tau}{\omega}.
\]
In Fig.~\ref{dispersion}, we show these dispersion relations calculated for the choice of the parameters $m, n, \tau, \nu$: $m=1.5m_e\ [\mathrm{g}],\ n= 2.5\times10^{22}\ [\mathrm{cm}^{-3}],\ \tau=1.0\times10^{-11}\ [\mathrm{s}^{-1}],\ \nu=3\times10^2 \ [\mathrm{cm^2s^{-1}}]$, which are typical values of the electron fluid in PdCoO${}_2$ in experiments~\cite{Moll2016, Mackenzie2017}. We note that the following results are qualitatively the same as for the parameters chosen for another hydrodynamic material WP${}_2$~\cite{Gooth2018}: $m, n, \tau, \nu$: $m=1.2m_e\ [\mathrm{g}],\ n= 2.9\times10^{21}\ [\mathrm{cm}^{-3}],\ \tau=5\times10^{-11}\ [\mathrm{s}^{-1}],\ \nu=3.8\times10^2 \ [\mathrm{cm^2s^{-1}}]$.

%%PdCoO2から持ってきていることを言及したい。

In particular, we can approximate the dispersion relations in the low-frequency limit $\omega\ll 1/\tau, \omega_p$ as, 
\[
\kk_1^2(\omega)\simeq k_0^2+ \frac{4\pi i \sigma_0\omega}{c^2},\ \ 
\kk_2^2(\omega)\simeq -\frac{1}{\nu \tau} -\frac{4\pi i\sigma_0\omega}{c^2}.
\]
Here we find that the right-hand side of the first equation corresponds to the low-frequency limit of the dispersion relation deduced from the Drude theory (See also Fig.~\ref{dispersion} (a) and (b)). This means that we can regard the mode $k_1(\omega)$ as the ``$Drude$-$like\  mode$'', and the mode $k_2(\omega)$ as the ``$hydrodynamic\ mode$''. 
In Fig.~\ref{dispersion} (c) and (d), we also show the $\nu$-dependence of the imaginary part of the dispersion relations. 
As seen from the figure, Im$[k_{1,2}(\omega)]$ becomes smaller as the viscosity becomes larger, implying that electromagnetic waves can penetrate more deeply as viscosity becomes larger. 
As seen in the following sections, the existence of two propagating modes and these dispersion relations play an important role in optical properties of electron fluids, such as the reflection and the transmission of electromagnetic waves.

\section{Reflectance \label{3}}

\begin{figure}[t]
\centering
\includegraphics[width =7cm]{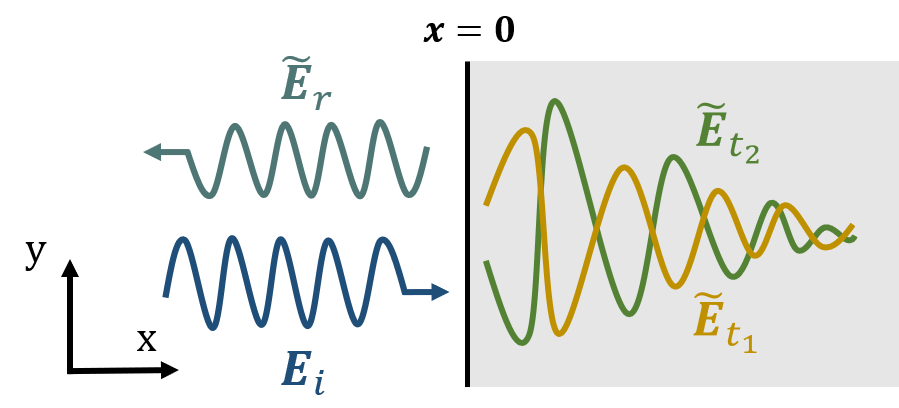}
\caption{Geometry of the reflection of vertically incident wave on the surface of 3D electron fluids. In electron fluids, there exist two propagating waves $\E_{t_1}$ and $\E_{t_2}$ corresponding to the dispersion relations $k_1(\omega)$ and $k_2(\omega)$.}
\label{setup}
\end{figure}

In this section, we consider the reflectance of 3D electron fluids. We suppose that the region $x>0$ is occupied with electron fluids (see Fig.~\ref{setup}). Here we deal with the case where the incident wave is linearly polarized vertically to the surface. For more general cases, see the Appendix. 

Under the above assumption, in the vacuum $(x<0)$, the AC fields are given by the sum of an incident wave $\E_i$ and an 
reflected wave $\E_r$ as follows: 
\begin{equation}\label{vacuum1}
\E =\E_i + \E_r, \  {\bm B} = {\bm B}_i+{\bm B}_r \ \ \ \ (x<0)
\end{equation}
\[
\E_i =\Re\left[ E_i \unit_y e^{ik_i x-i\omega t}\right],\ 
 {\bm B}_i=\Re\left[ k_i'E_i \unit_z e^{ik_i x -i\omega t}\right],
\]
\[
\E_r = \Re\left[ \tilde{E}_r \unit_y e^{ik_r x-i\omega t} \right],\ 
{\bm B}_r= \Re\left[ k_r'\tilde{E}_r \unit_z e^{ik_r x -i\omega t}\right],
\]
where $k_{i}=-k_{r}$ and we introduce a dimensionless parameter $k' \equiv ck/\omega$. 
$\tilde{E}_{r}$ is a complex parameter to be determined by imposing  appropriate boundary conditions. 

On the other hand, in electron fluids $(x>0)$, the transmitted wave $\E_t$ is composed of two  propagating modes $\E_{t_1}$ and $\E_{t_2}$ corresponding to the dispersion relations $k_1(\omega)$ and $k_2(\omega)$ as follows: 
\[
\E_{t} = \E_{t_1}+\E_{t_2}, \  {\bm B} = {\bm B}_{t_1}+{\bm B}_{t_2} \ \ \ \ (x>0)
\]
\[
\E_{t_1} = \Re\left[ \tilde{E}_{t_1} \unit_y e^{i{\tilde{k}}_{1} x-i\omega t} \right],\ 
{\bm B}_{t_1}=\Re\left[ \tilde{k}_{1}'\tilde{E}_{t_1} \unit_z e^{i{\tilde{k}}_{1} x -i\omega t}\right],
\]
\[
\E_{t_2} =\Re\left[ \tilde{E}_{t_2} \unit_y e^{i{\tilde{k}}_{2} x-i\omega t} \right],\ 
{\bm B}_{t_2}=\Re\left[ \tilde{k}_{2}'\tilde{E}_{t_2} \unit_z e^{i{\tilde{k}}_{2} x -i\omega t}\right],
\]
where we have added tildes to $k_1$ and $k_2$ to manifest that these variables are complex numbers. $\tilde{E}_{t_1}$ and $\tilde{E}_{t_2}$ are also complex parameters to be determined by imposing  appropriate boundary conditions. 

In the hydrodynamic regime, as just described, we have three undetermined parameters in total, while the conventional theory contains only two boundary conditions, that is, continuity conditions of electric and magnetic fields, 
\begin{equation}\label{electric BC}
E_y(x=+0)=E_y(x=-0)\ \Leftrightarrow\ E_{i}+\tilde{E}_{r}=\tilde{E}_{t_1}+\tilde{E}_{t_2},
\end{equation}
\begin{equation}\label{magnetic BC}
B_z(x=+0)=B_z(x=-0)\ \Leftrightarrow\  B_i+\tilde{B}_r=\tilde{B}_{t_1}+\tilde{B}_{t_2}.
\end{equation}
In general, when spatial dispersion exists and the number of propagating modes increases to more than one, the conventional boundary conditions are insufficient to describe the connections of electromagnetic fields at the surface. 
This is called the additional boundary condition (ABC) problem~\cite{Pekar1957, Halevi}. In our cases, the appropriate ABC is the no-slip BC at the surface, that is, we impose the following condition on electron fluids: 
\begin{equation}\label{no-slip BC}
v_y(x=+0)=0.
\end{equation}

Summarizing, we have to determine the parameters $\tilde{E}_{r}$, $\tilde{E}_{t_1}$ and $\tilde{E}_{t_2}$ by solving the simultaneous equations derived from boundary conditions (\ref{electric BC}), (\ref{magnetic BC}) and (\ref{no-slip BC}). 
This procedure can be done readily, and as a result, we obtain the following expression of the reflectance:  
\begin{equation}
R = \left|\frac{E_r}{E_i}\right|^2= \left| \frac{(a-b)k_i-(\tilde{k}_{t_1}a-\tilde{k}_{t_2}b)}{(a-b)k_i+(\tilde{k}_{t_1}a-\tilde{k}_{t_2}b)} \right|^2, 
\end{equation}
where we have introduced the parameters $a$ and $b$, which are defined so as to satisfy the following relation: 
\[
 \frac ab = \frac{1-i\omega \tau + \nu \tau \tilde{k}_{t_1}^2}{1-i\omega \tau + \nu \tau \tilde{k}_{t_2}^2}.
\]

\begin{figure}[t]
\centering
\includegraphics[width =8.5cm]{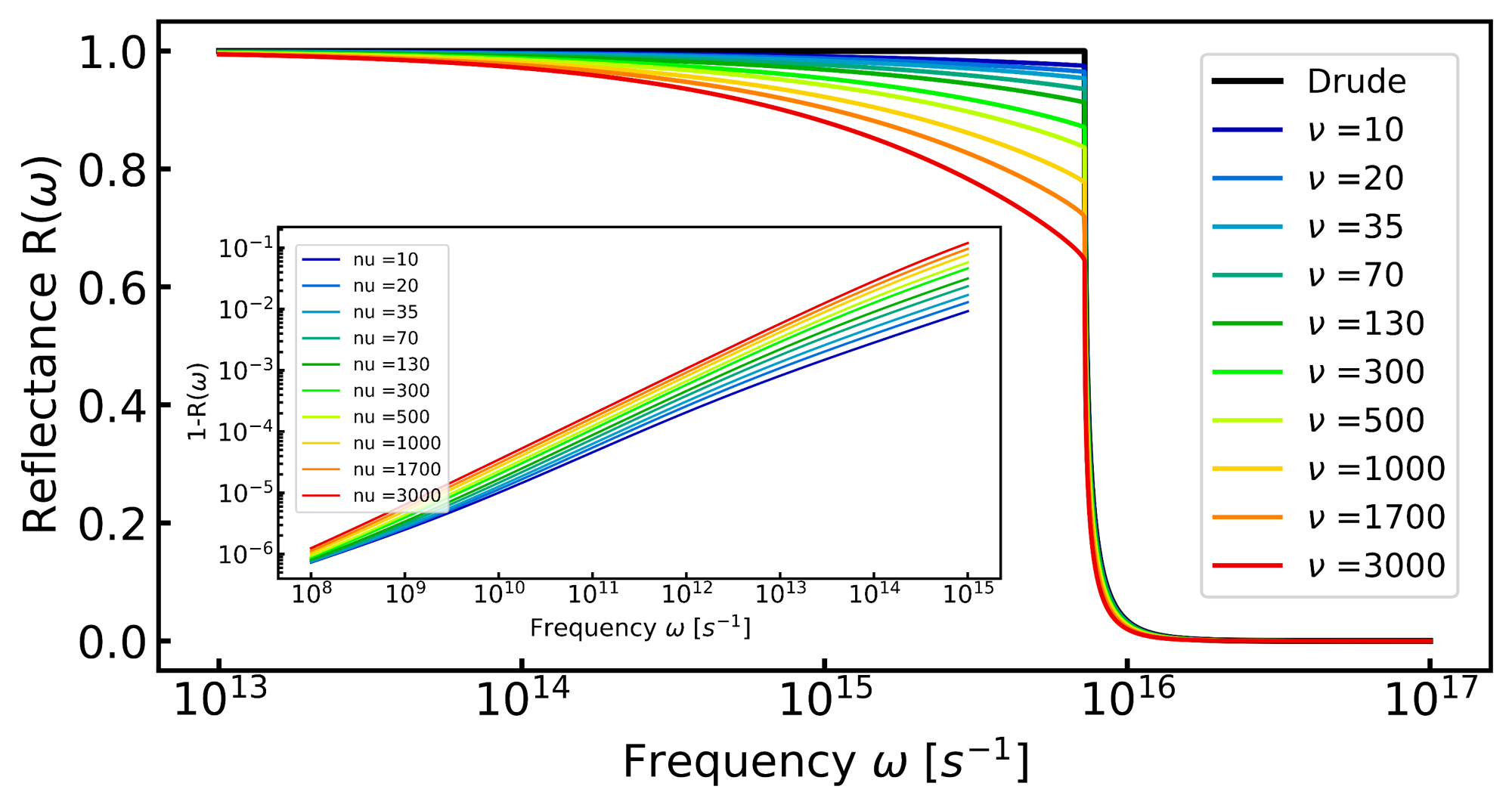}
\caption{The reflectance spectra of 3D electron fluids with various values of the viscosity. In this plot, we set the parameter other than viscosity $\nu$ to the typical values in PdCoO${}_2$ and, for comparison, show the estimation by the Drude theory (black line). As is well known, the spectra exhibit a sharp drop at the plasma frequency $\omega_p\simeq 7\times 10^{15}$. 
The inset shows the deviation of reflectance from 1 in the low-frequency regime. }
\label{reflectance}
\end{figure}

For comparison, in Fig.~\ref{reflectance}, we show the reflectance spectra of 3D electron fluids with various values of the viscosity, together with the estimation by the Drude theory. 
As is well known, each spectrum exhibits a sharp drop around the plasma frequency $\omega_p\simeq 7\times 10^{15}$. It can be seen that, as viscosity becomes larger, the reflectivity deviates from 1 more prominently, especially in the vicinity of the plasma frequency $\omega_p$. To apply the hydrodynamic theory to electrons, however, the system needs to satisfy the condition $\tau_{ee}\ll 1/\omega$, as mentioned in Sec.~\ref{introduction}. In such a low-frequency regime, since typical values of $1/\tau_{ee}$ in hydrodynamic materials known at present are much smaller than $\omega_p$, the deviation becomes relatively small and so it seems to be difficult to observe these hydrodynamic effects through experiments in typical materials. For this reason, the reflectance measurement may not be suitable for an experimental probe of hydrodynamic effects. However there is still a possibility to observe these effects if the hydrodynamic regime is realized in materials which show a much smaller value of  $\tau_{ee}$ and a lower carrier density $n\ (\propto \omega_p^2)$ than hydrodynamic materials known at present.

\section{Transmittance through thin ultrapure metals \label{4}}

\begin{figure}[t]
\centering
\includegraphics[width =7cm]{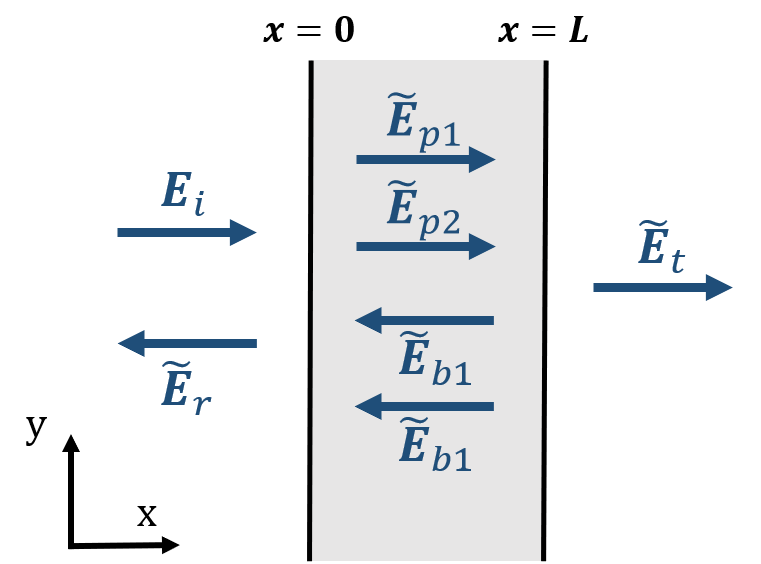}
\caption{Sketch of the transmission of electronmagnetic waves through 3D electron fluids. $\tilde{E}_{p_{f1}}$ ($\tilde{E}_{p_{b1}}$) and $\tilde{E}_{p_{f2}}$ ($\tilde{E}_{p_{b2}}$) respectively denote the complex amplitudes of forward (back) propagating waves corresponding to the dispersion relation $\kk_1(\omega)$ and $\kk_2(\omega)$.}
\label{setup2}
\end{figure}

\begin{figure*}[t]
  \centering
  \begin{tabular}{ll}
(a)  &(b) \\
   \includegraphics[width=7.5cm]{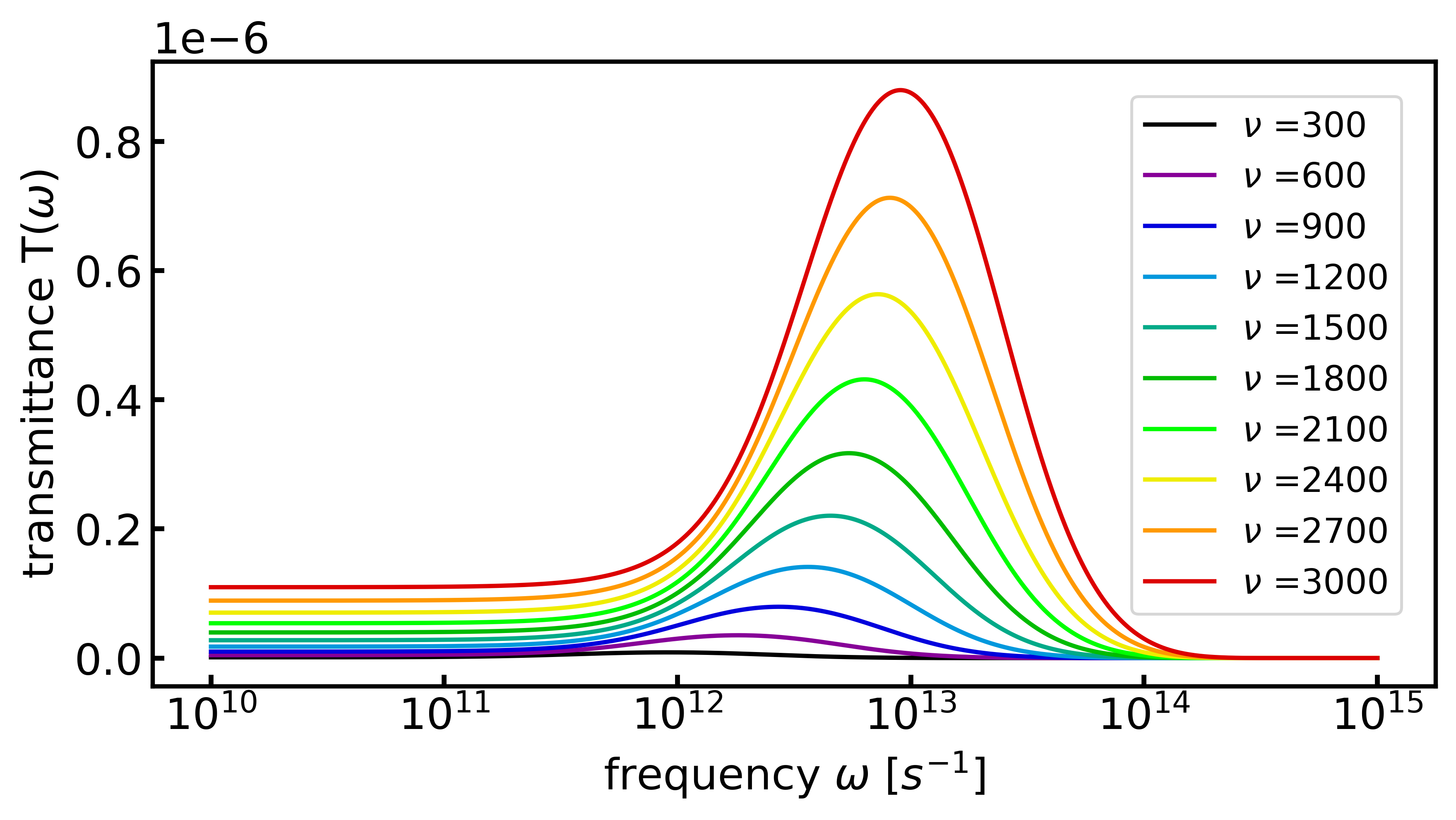}&
    \includegraphics[width=7.5cm]{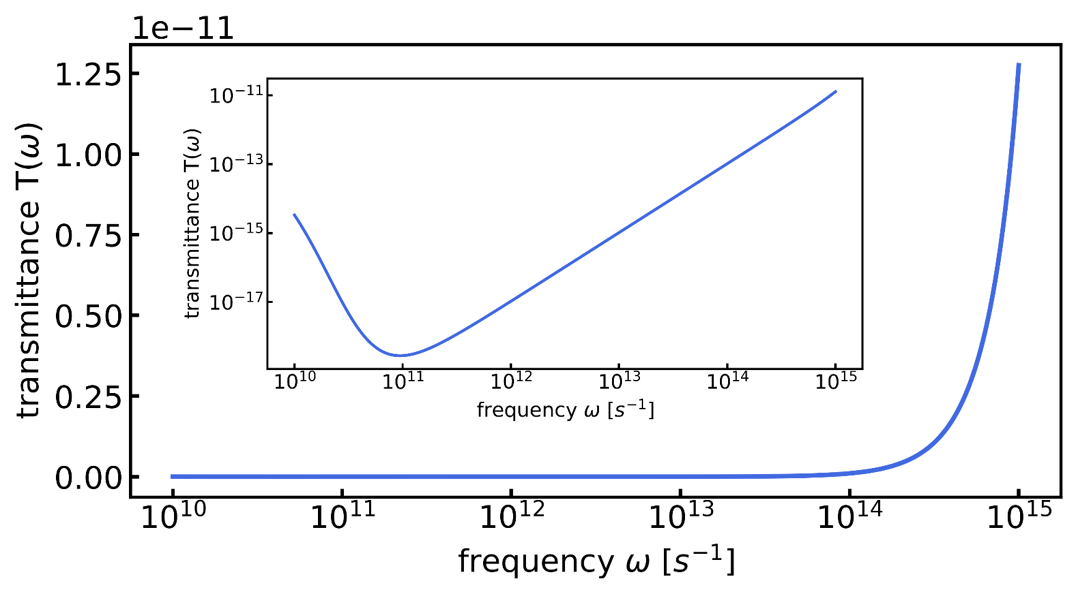}\\
(c) & (d) \\
   \includegraphics[width=7.5cm]{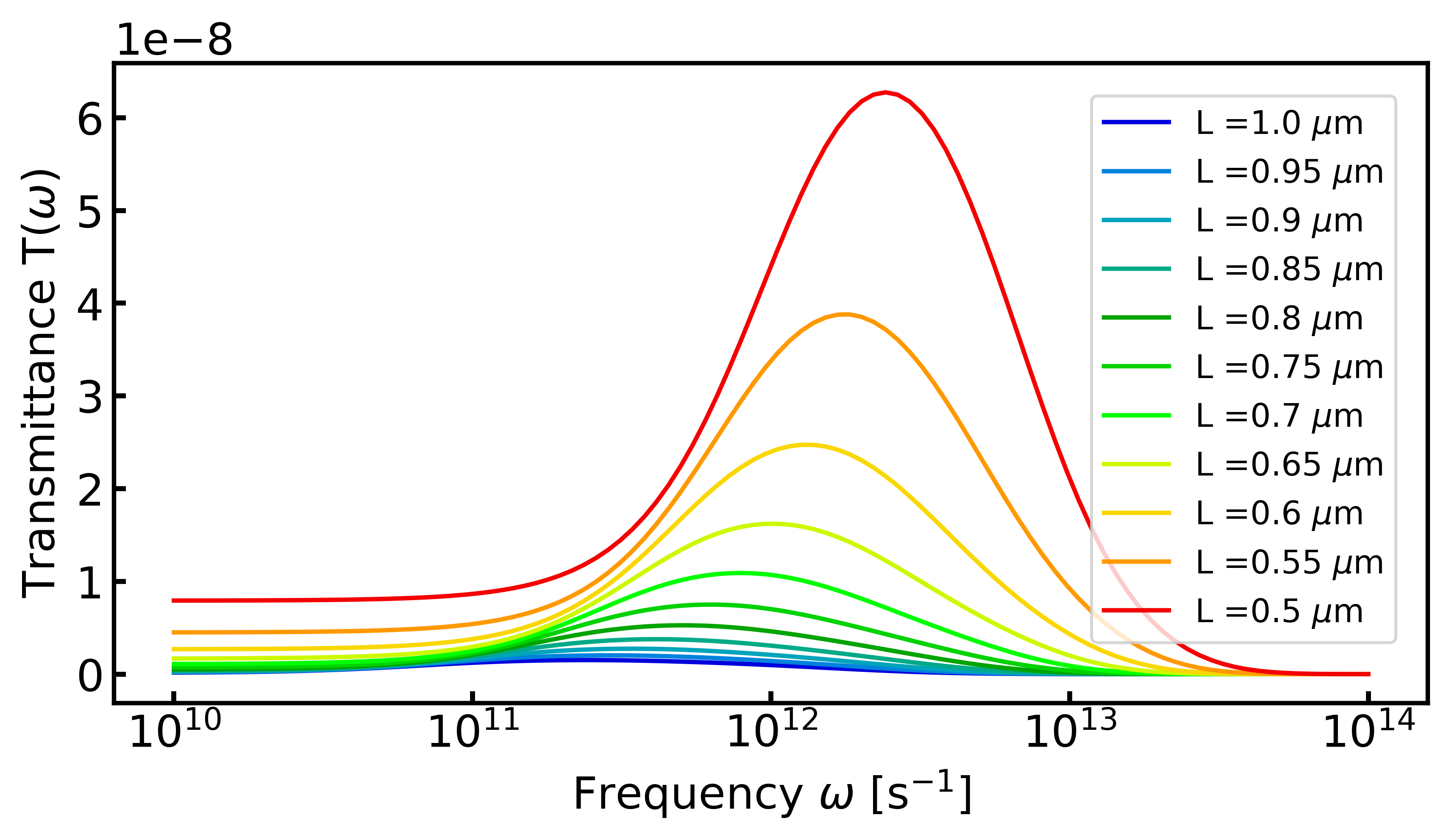}&
    \includegraphics[width=7.5cm]{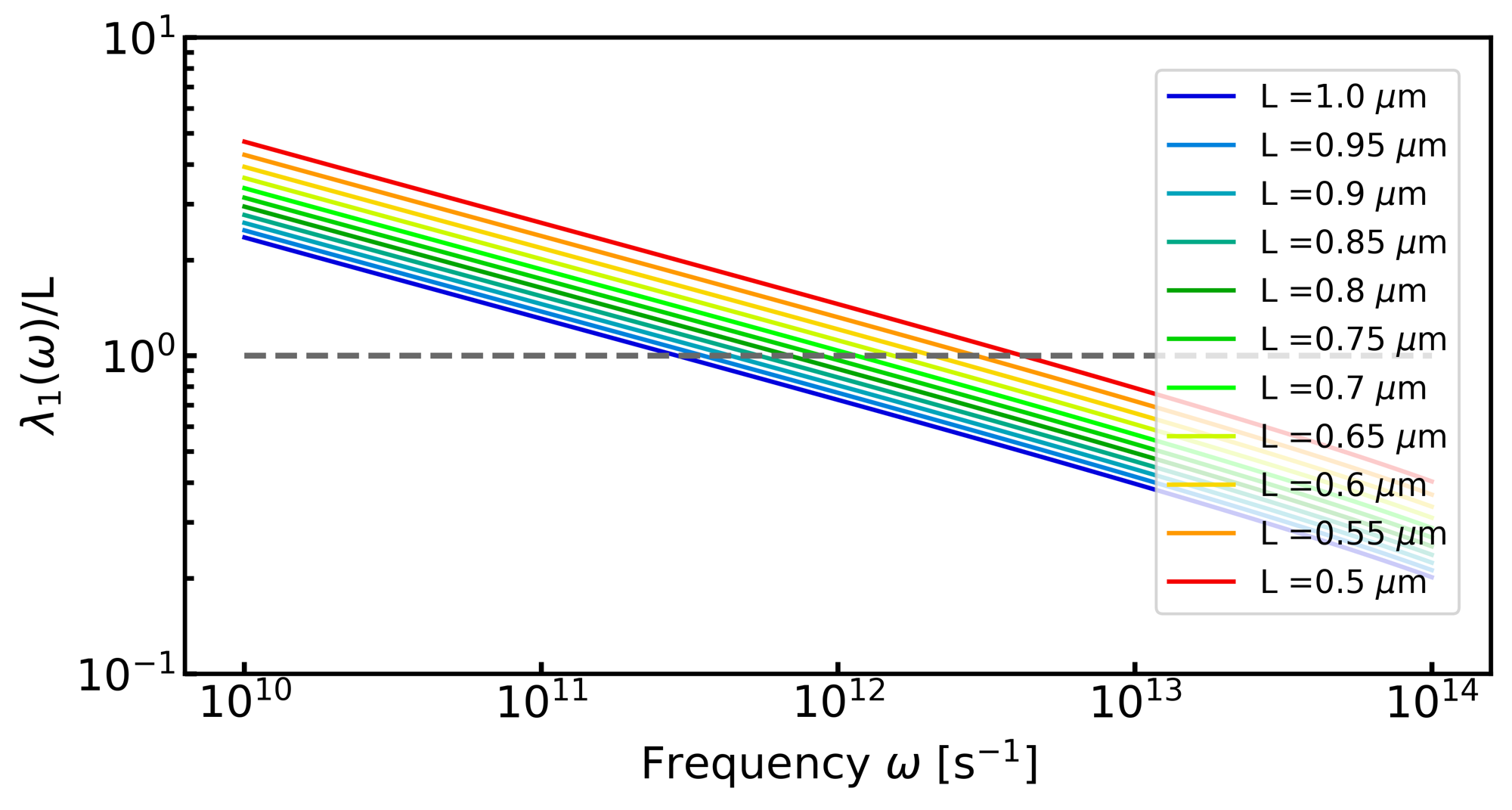}
  \end{tabular}
\caption{(a)(b) Transmittance spectra through a 0.5$\mu$m thin metal in the hydrodynamic regime (a) and in the Drude regime (b): In the hydrodynamic regime, the amplitudes of transmittance are five or more digits greater than that in the Drude regime and the spectra show a characteristic peak structure. (c) Transmittance spectra through thin metals with various choices of thickness (viscosity $\nu$ is fixed at $3\times10^2 \ [\mathrm{cm^2s^{-1}}]$). (d) Ratio of the width of thin metals $L$ and wavelength of the $Drude$-$like$ $mode$ $\lambda_1(\omega)$: Comparing with (c), we can understand that the peak appears at the frequency where $\lambda_1(\omega)$ becomes equal to $L$ (gray dotted line). }
  \label{transmittance}
\end{figure*}

Next, we consider the transmission of electromagnetic waves through a thin ultrapure metal. 
As described in Appendix, even in the hydrodynamic regime, the Snell's law of refraction is valid for each propagating mode. Therefore, when the sample has a prismatic structure, an incident monochromatic wave is separated into two directions due to the difference of the refractive indices of these modes. Moreover, since the transmittance reflects the dispersion relations of the metal, we may be able to determine the viscosity from the observed transmittance spectrum. For these reasons, such a transmission phenomenon is naively expected to be an efficient experimental probe of hydrodynamic effects. However, to ensure the validity of the hydrodynamic theory, we need to make a metallic slab thicker than $l_{ee}$, which is typically in a micrometer range. In terms of the Drude theory, the transmittance through metals of such thickness seems to be too small to be measured experimentally. 

In what follows, in response to the above discussion, we estimate the transmittance through thin metals in the hydrodynamic regime and demonstrate that it becomes much larger than in the Drude regime because of the viscosity effect and, as a result, we can detect experimentally the hydrodynamic effect through the transmittance. 

Let us now consider for simplicity the transmittance through a slab with thickness of $L$ (see Fig.~\ref{setup2}). Here, as with the previous section, we deal with the case where an incident wave is linearly polarized and directed vertically to the surface. In the vacuum, the ac field is given by the same equation~(\ref{1}) as in Sec.~\ref{3} in $x<0$, and, in $x>L$, 
\[
\E = \E_t,\ {\bm B} = {\bm B}_t \ \ \ \ (x>L),
\]
\[
\E_t =\Re\left[ \tilde{E}_t \unit_y e^{ik_i (x-L)-i\omega t}\right],\ 
 {\bm B}_t=\Re\left[ k_i' \tilde{E}_t \unit_z e^{ik_i (x-L) -i\omega t}\right],
\]
where $\tilde{E}_t$ is a complex parameter to be determined. 
In electron fluids ($0<x<L$), we need to consider a forward propagating wave and a backward propagating wave for each dispersion branch and as a result the ac electric field is described as follows: 
\[
\E = \E_{p_{f1}} + \E_{p_{f2}} + \E_{p_{b1}} + \E_{p_{b2}},
\]   
\[
\E_{p_{f1}} =\Re\left[ \tilde{E}_{p_{f1}} \unit_y e^{i\tilde{k}_{f1}x-i\omega t}\right], 
\]\[
\E_{p_{f2}} =\Re\left[ \tilde{E}_{p_{f2}} \unit_y e^{i\tilde{k}_{f2} x-i\omega t}\right],
\]
\[
\E_{p_{b1}} =\Re\left[ \tilde{E}_{p_{b1}} \unit_y e^{i\tilde{k}_{b1}x-i\omega t}\right], 
\]\[
\E_{p_{b2}} =\Re\left[ \tilde{E}_{p_{b2}} \unit_y e^{i\tilde{k}_{b2} x-i\omega t}\right],
\]
where $\tilde{E}_{p_{f1}},\  \tilde{E}_{p_{f2}},\ \tilde{E}_{p_{b1}}$, and $\tilde{E}_{p_{b2}}$ are complex parameters to be determined and $\tilde{k}_{f1}=-\tilde{k}_{b1}=\tilde{k}_1(\omega)$, $\tilde{k}_{f2}=-\tilde{k}_{b2}=\tilde{k}_2(\omega) $. 
The AC magnetic fields are also decribed in a similar form. Moreover, the velocity field of the fluids is described by the sum of the velocity fields corresponding to each mode, 
\[
\vv = \sum_\alpha \vv_\alpha,
\]
where the sum runs over $\alpha= f1,f2,b1,b2$, and $\vv_\alpha$ is given in terms of the mobility $\mu_\alpha$ as follows: 
\[
\vv_\alpha = \mu_\alpha \E_\alpha,\ \ \mu_\alpha \equiv \frac{\sigma_\perp(\tilde{\kk}_\alpha, \omega)}{(-e)n}.
\]

\begin{figure*}[t]
\centering
\includegraphics[width =13cm]{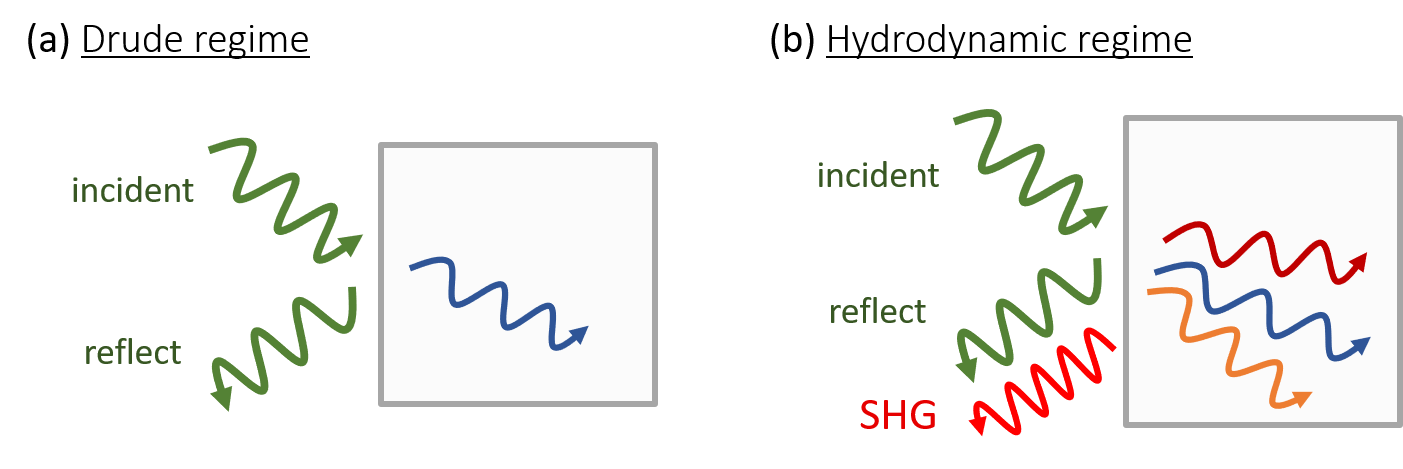}
\caption{Sketch of the mechanism of the second-order harmonic generation (SHG) in the hydrodynamic picture. In the hydrodynamic regime, an incident plane wave drives three propagating modes (two transverse modes and one longitudinal mode) in electron fluids and they couple each other nonlocally through the nonlinear terms in the hydrodynamic equation, which leads to nonlocal nonlinear optical responses, such as SHG.  }
\label{SHG}
\end{figure*}

As described in the previous section, we can determine the above parameters by imposing, at the surface ($x=0,L$), conditions of continuity for the electromagnetic fields,  
\begin{equation}
E_y(+0)=E_y(-0)\ \Leftrightarrow\ E_{i}+\tilde{E}_{r}=\sum_{\alpha}\tilde{E}_{p_\alpha},
\end{equation}
\begin{equation}
B_z(+0)=B_z(-0)\ \Leftrightarrow\ E_{i}-\tilde{E}_{r}=\sum_{\alpha} \tilde{k}_\alpha' \tilde{E}_{p_\alpha},
\end{equation}
\begin{equation}
E_y(L+)=E_y(L-)\ \Leftrightarrow\  \tilde{E}_t =\sum_{\alpha}\tilde{E}_{p_\alpha}e^{i\tilde{k}_\alpha L},
\end{equation}
\begin{equation}
B_z(L+)=B_z(L-)\ \Leftrightarrow\  \tilde{E}_t=\sum_{\alpha} \tilde{k}_\alpha' \tilde{E}_{p_\alpha}e^{i\tilde{k}_\alpha L},
\end{equation}
and the no-slip BC on the velocity field, 
\begin{equation}
v_y(x=+0)=0\ \Leftrightarrow\  \sum_\alpha \mu_\alpha  \tilde{E}_\alpha =0,
\end{equation}
\begin{equation}
v_y(x=L-)=0\ \Leftrightarrow\ \sum_\alpha \mu_\alpha  \tilde{E}_\alpha e^{i\tilde{k}_\alpha L}=0.
\end{equation}
We can easily solve these simultaneous equations for $\tilde{E}_r, \tilde{E}_t, \tilde{E}_\alpha$ and finally obtain the transmittance $T =\left|{\tilde{E}_t}/{E_i}\right|^2$.

In Fig.~\ref{transmittance}(a), we show the viscosity dependence of the transmittance spectra through 3D electron fluids of $0.5\ \mu$m thickness. We have calculated these results for the choice of the same parameters as in Sec.~\ref{2}. Compared with the estimation from the Drude theory (Fig.~\ref{transmittance}(b)), the amplitudes of transmittance in the hydrodynamic regime are $10^5$ or more times larger than that in the Drude regime and the spectra show a characteristic peak structure. Therefore, we reach the remarkable conclusion that, in the hydrodynamic regime, the transmittance becomes large enough to observe experimentally, even through the metals of $\mu$m-order thickness. Especially in low-frequency regime, the Drude-like mode seems to largely contribute to the transmittance, since the imaginary part of the wavenumber, which is  inversely proportional to the damping length, becomes relatively small as seen in Fig.~\ref{dispersion}(b). 

To reveal the origin of the peak of transmittance spectra, we show the thickness dependence of transmittance spectra in Fig.~\ref{transmittance}(c) and the ratio of the thickness and the wavelength of the Drude-like mode $\lambda_1(\omega)\equiv 2\pi/{\rm Re}[k_1(\omega)]$ in Fig.~\ref{transmittance}(d), where we have fixed the viscosity at $3\times10^2 \ [\mathrm{cm^2s^{-1}}]$. By comparison of these figures, we can understand that the peak frequency corresponds to the resonance frequency where the wavelength of the Drude-like mode becomes equal to the thickness of the metals. This means that identifying the peak frequency in transmittance spectrum experimentally leads to the dispersion relation of the Drude-like modes, which include the information of the viscosity of electron fluids as in Eq.~(\ref{bunsan}).

\section{summary and discussion \label{5}}

In summary, we have developed a basic framework of optical responses in the hydrodynamic regime. In particular, we have revealed the hydrodynamic effects on optical linear responses in 3D hydrodynamic metals and quantitative signatures of these effects on optical observables, i.e. reflectance and transmittance, comparing with the Drude theory. In the hydrodynamic regime, two propagating modes emerge due to the nonlocality of electron fluids and lead to the change of the behavior in reflection and transmission phenomena from those estimated by the Drude theory, which neglects electron-electron scatterings and the nonlocality. We have shown that, in regard to hydrodynamic materials known at present, it may be difficult to detect hydrodynamic effects from the measurement of the reflectance because the reflectance modulation is relatively small in the low-frequency regime where hydrodynamic theory is applicable to the electron dynamics. 
On the other hand, the transmittance spectrum has an extremely large value compared with that in the Drude regime and a characteristic peak structure corresponding to the resonance of the ``Drude-like'' mode. 
These remarkable facts have lead us to conclude that the transmittance measurement is an efficient method to evidence the existence of a hydrodynamic regime and determine the viscosity of electron fluids. 

Although, in this paper, we focused on the linear optical responses of electron fluids, the existence of several propagating modes opens up a new possibility of nonlinear optical responses in the hydrodynamic regime. 
In general, centrosymmetric materials, such as PdCoO${}_2$, have no (local) second-order nonlinear susceptibility in bulk due to the constraint of the symmetry. In the hydrodynamic regime, however, as described in Appendix, an incident plane wave drives three propagating modes (two transverse modes and one longitudinal mode) in electron fluids and they couple each other nonlocally through the nonlinear terms  in the hydrodynamic equation~(\ref{general}) (see Fig.~\ref{SHG}). 
This mechanism leads to the finite nonlocal second-order nonlinear susceptibility of 3D electron fluids. 
This contribution to nonlocal nonlinear responses seems to characterize the optical response of 3D electron fluids and gives the evidence of the multi-branch structure of the transverse modes. 

We note, however, that there is an overlooked problem in the above discussion. Although, in the second-order responses, we need to deal with not only transverse modes, but also a longitudinal mode, the damping length of a longitudinal wave is of a nanometer scale for typical values of parameters of bulk metallic materials, which is so small that the hydrodynamic theory is no longer applicable to describe the electron dynamics and, to perform a reliable calculation, we need to go back to more microscopic descriptions than the hydrodynamic theory. Nevertheless, the above discussion may be meaningful if the hydrodynamic regime is realized in some dilute metals at high temperature, where
$\tau_{ee}$ is expected to be much smaller and the damping length, which is inversely proportional to the carrier density $n$ in the low-frequency limit, becomes larger than that for hydrodynamic materials observed so far. This issue will be an interesting future work bridging the areas of electron hydrodynamics and nonlinear optics.

\section{acknowledgments}

We are thankful to Hikaru Watanabe and Akito Daido for valuable discussions. We also thank Koichiro Tanaka for providing helpful comments from an experimental point of view. K.T. acknowledges helpful discussions with Thomas Scaffidi and Joel E. Moore. This work is supported by a Grant-in-Aid for Scientific Research on Innovative Areas Topological Materials Science (KAKENHI Grant No. JP15H05855) and also JSPS KAKENHI (Grants JP16J05078, JP18H01140 and JP19H01838). K.T. thanks JSPS for support from Research Fellowship for Young Scientists and Overseas Research Fellowship.

\vspace{1cm}

---$\bm{N}\bm{o}\bm{t}\bm{e}$ $\bm{a}\bm{d}\bm{d}\bm{e}\bm{d}$---
After uploading this preprint, we have noticed that the results similar to ours had already been obtained by D. Forcella, J. Zaanen, D. Valentinis and D. van der Marel~\cite{Forcella2014}.

\appendix
\section{Reflectance for a diagonally oriented incidence wave}

\begin{figure}[t]
\centering
\includegraphics[width =8cm]{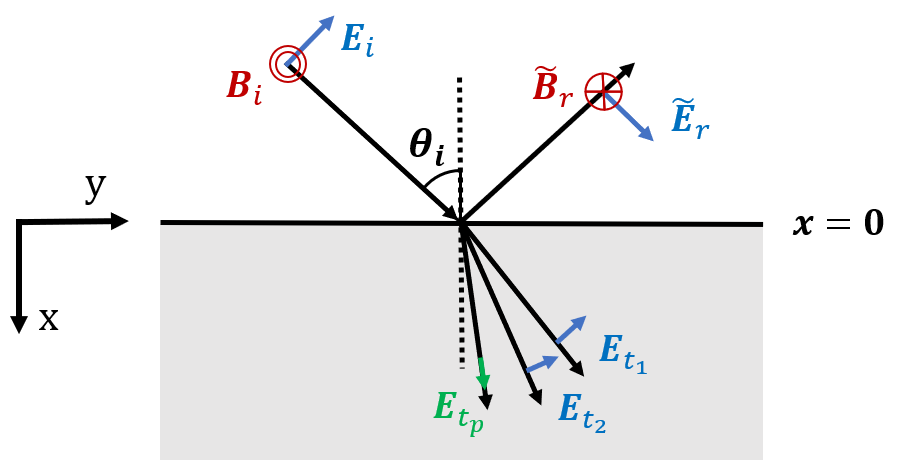}
\caption{Geometry of p-polarized light scattering on the surface of 3D electron fluids. }
\label{setup3}
\end{figure}

In this Appendix, we consider the reflectance of 3D electron fluids for an incident wave oriented vertically to the surface. We suppose that the region $x>0$ is occupied with electron fluids (see Fig.~\ref{setup3}). 
Here, to make the following discussion applicable to layered systems, we assume that the incident electromagnetic wave is $p$-polarized and the electric field $\E_i(\rr,t)$ is in the $x$-$y$ plane,
\[
\E_i(\rr,t) = \Re\left[{\bm {A}}_i e^{i\kk_i\cdot \rr-i\omega t} \right] .
\]
\[
\B_i(\rr,t) = \Re\left[\frac{c\kk_i\times {\bm{A}}_i}{\omega} e^{i\kk_i\cdot \rr-i\omega t} \right] .
\]
where ${\bm A}_i$ and $\kk_i$ are vectors parallel to the $x$-$y$ plane and orthogonal to each other. In the hydrodynamic regime, the reflected waves $\E_r,\B_r$ are described as
\[
\E_r =  \Re\left[ {\bm \tA}_r e^{i\kk_r\cdot \rr-i\omega t}\right],\ 
{\bm B}_r=  \Re\left[ \frac{c\kk_r\times {\bm \tA}_r}{\omega} e^{i\kk_r \cdot \rr -i\omega t}\right],
\]
and the transmitted waves $\E_t,\B_t$ are described by the sum of two transverse modes and one longitudinal mode,
\[
\E_t = \E_{t_1}+\E_{t_2}+\E_{t_p} ,\   \B_t = {\bm B}_{t_1} + {\bm B}_{t_2} + {\bm B}_{t_p},
\]
\[
\E_{t_1} = \Re\left[ {\bm \tA}_{t_1} e^{i{\tilde{\kk}}_{1}\cdot \rr-i\omega t}\right],\ 
{\bm B}_{t_1}= \Re\left[ \frac{c\tilde{\kk}_{1}\times {\bm \tA}_{t_1}}{\omega} e^{i\tilde{\kk}_{1} \cdot \rr -i\omega t}\right],
\]
\[
\E_{t_2} = \Re\left[ {\bm \tA}_{t_2} e^{i{\tilde{\kk}}_{2}\cdot \rr-i\omega t}\right],\ 
{\bm B}_{t_2}= \Re\left[ \frac{c{\tilde{\kk}}_{2}\times {\bm \tA}_{t_2}}{\omega} e^{i\tilde{\kk}_{2} \cdot \rr -i\omega t}\right],
\]
\[
\E_{t_p} = \Re\left[ {\bm \tA}_{t_p} e^{i\tilde{\kk}_{p}\cdot \rr-i\omega t}\right],\ 
{\bm B}_{t_p}= {\bm 0},
\]
where ${\tilde{\kk}}_{1}(\omega)$, ${\tilde{\kk}}_{2}(\omega)$, and ${\tilde{\kk}}_{p}(\omega)$ are the complex wavenumbers corresponding to dispersion relations introduced in Sec.~\ref{2}. As seen below, to satisfy the boundary condition, we need to consider the contribution of a longitudinal mode to the transmitted wave, which is also a characteristic feature of the optical responses in the hydrodynamic regime. 
Imposing the continuity of electromagnetic fields, we obtain the equation for these wavenumbers as follows:
\[
\kk_i\cdot {\bm t}=\kk_r\cdot {\bm t}=\tilde{\kk}_{1}\cdot {\bm t}=\tilde{\kk}_{2}\cdot {\bm t}=\tilde{\kk}_{p}\cdot {\bm t}.
\]
As can be easily understood, this leads to the reflection law
\begin{equation}
k_{ix}=-k_{rx},\  k_{iy}=k_{ry},
\end{equation}
and Snell's law
\begin{equation}\label{Snell}
\sin \theta_i = \frac{c}{\omega} |\Re \tilde{\kk}_\alpha| \sin \theta_\alpha  \ \  (\alpha=1,2,p)
\end{equation}
where $\theta_i$ and $\theta_\alpha$ are the angle of $\kk_i$ and $\Re[\kk_\alpha]$ measured from the normal of the surface and, for simplicity, we neglect the background refractive index. Using Eq.~(\ref{Snell}), we can rewrite the transverse wave condition for $\kk_{1,2}$ and the longitudinal wave condition $\kk_{p}$ as follows:
\begin{equation}\label{yoko}
\tilde{\kk}_{\alpha} \cdot {\bm \tA}_{\alpha}=0 \ \Leftrightarrow \ \frac{\tA_{\alpha,x}}{\tA_{\alpha,y}}=- \frac{\tilde{k}_{\alpha y}}{\tilde{k}_{\alpha x}}=  -\tan \tilde{\theta}_\alpha \ \  (\alpha=1,2)
\end{equation}
\begin{equation}\label{tate}
\tilde{\kk}_t\times {\bm \tA}_{p}=0\  \Leftrightarrow\ \frac{\tA_{p,x}}{\tA_{p,y}}= \frac{\tilde{k}_{px}}{\tilde{k}_{py}}= \frac{\cos\tilde{\theta}_t}{\sin\tilde{\theta}_p}= \cot \tilde{\theta}_p,
\end{equation}
where we have defined complex angle $\tilde{\theta}_\alpha$ for each mode as 
\begin{equation}
\sin \theta_i = \tilde{n_\alpha}\sin \tilde{\theta}_\alpha,\ \  \tilde{n}_\alpha\equiv\sqrt{\frac{{\bm \tilde{\kk}}_t^2 }{k_0^2} } \ \ (\alpha = 1,2,p).
\end{equation}

Next, we consider the no-slip BC for the velocity field at surface. The velocity field of the fluids is described by the sum of the velocity fields corresponding to each mode,
\[
\vv = \sum_{\alpha=1,2,p} \vv_\alpha,
\]
where $\vv_\alpha$ is given in terms of the mobility $\mu_\alpha$ as follows: 
\[
\vv_\alpha = \mu_\alpha \E_{t_\alpha},\ \ \mu_\alpha \equiv \frac{\sigma_\perp(\tilde{\kk}_\alpha, \omega)}{(-e)n},\ \ (\alpha=1,2)
\] 
\[
\vv_p = \mu_p \E_{t_p},\ \ \mu_p \equiv \frac{\sigma_\perp(\tilde{\kk}_p, \omega)}{(-e)n}.
\] 
Imposing the no-slip BC ($v_x=0$, $v_y=0$) on the velocity field, we obtain simultaneous equations for $ {\bm \tA}_{t_1}$, ${\bm \tA}_{t_2}$, and ${\bm \tA}_{t_p}$ and, solving these equations together with Eq.~(\ref{yoko},~\ref{tate}), we reach the following relations: 
\begin{equation}
\begin{split}
		\frac{\tA_{t_2,y}}{\tA_{t_1,y}}&= -\frac{\mu_1}{\mu_2}\frac{\tan\tilde{\theta}_{1} +\cot \tilde{\theta}_{p}}
		{\tan\tilde{\theta}_{2} +\cot \tilde{\theta}_{p}}\\
		&=-\frac{1-i\omega t +\nu \tau \tilde{\kk}_{2}^2}{1-i\omega t +\nu \tau \tilde{\kk}_{1}^2}\cdot  \frac{\tan\tilde{\theta}_{1} +\cot \tilde{\theta}_{p}}
		{\tan\tilde{\theta}_{2} +\cot \tilde{\theta}_{p}}.
\end{split}
	\end{equation}
	\begin{equation}
\begin{split}
	\frac{\tA_{t_p,y}}{\tA_{t_1,y}}&= \frac{\mu_1}{\mu_p}\frac{\tan\tilde{\theta}_{1} -\tan \tilde{\theta}_{2}}
	{\tan\tilde{\theta}_{2} +\cot \tilde{\theta}_{p}}\\
	&=\frac{1-i\omega t +\left(
	\frac {iK}{mn_0\omega} +\nu +\gamma
	\right) \tau \tilde{\kk}_{p}^2}{1-i\omega t +\nu \tau \tilde{\kk}_{1}^2}\cdot
	\frac{\tan\tilde{\theta}_{1} -\tan \tilde{\theta}_{2}}
	{\tan\tilde{\theta}_{2} +\cot \tilde{\theta}_{p}}.
\end{split}
\end{equation}

Finally, imposing the continuous conditions on electric and magnetic field, we obtain the reflection coefficient and the transmission coefficient as follows: 
\begin{equation}\label{reflection coefficient}
		\frac{A_{t_1y}}{A_i} = \frac{2\cos\theta_i}{\left(1+\frac{\tA_{t_2,y}}{\tA_{t_1,y}} +\frac{\tA_{t_p,y}}{\tA_{t_1,y}} \right)
		+\cos\theta_i \left(
		\frac{\tilde{n}_{1}}{\cos\tilde{\theta}_{1}}  +
		\frac{\tilde{n}_{2}}{\cos\tilde{\theta}_{t_2}}   \frac{\tA_{t_2,y}}{\tA_{t_1,y}}
		\right),
		}
	\end{equation}
	\begin{equation}
		\frac{A_r}{A_i} = -1 + \frac{2  \left(1+\frac{\tA_{t_2,y}}{\tA_{t_1,y}} +\frac{\tA_{t_p,y}}{\tA_{t_1,y}}   \right)}{ \left(1+\frac{\tA_{t_2,y}}{\tA_{t_1,y}} +\frac{\tA_{t_p,y}}{\tA_{t_1,y}} \right)  + \cos\theta_i \left(
		\frac{\tilde{n}_{1}}{\cos\tilde{\theta}_{1}}  +
		\frac{\tilde{n}_{2}}{\cos\tilde{\theta}_{2}}   \frac{\tA_{t_2,y}}{\tA_{t_1,y}}
		\right) }.
	\end{equation}
The reflectance of electron fluids is obtained for any incident angle $\theta_i$ and frequency $\omega$ by calculating the square of absolute value of reflection coefficient~(\ref{reflection coefficient}).

\bibliographystyle{apsrev4-1}
\bibliography{ref}

\end{document}